\documentclass[11pt,a4paper]{article}
\usepackage{fix-cm}
\usepackage[T1]{fontenc}
\usepackage{lmodern}
\usepackage{amsmath,amssymb,booktabs,tabularx,array,graphicx}
\usepackage{tikz}
\usetikzlibrary{arrows.meta,positioning,fit,calc,backgrounds}

\usepackage[a4paper,margin=24mm,headheight=14pt,headsep=8mm]{geometry}
\usepackage[expansion=false]{microtype}
\usepackage{caption}
\usepackage{enumitem}
\setlist{itemsep=2pt,topsep=4pt}
\usepackage{fancyhdr}
\usepackage{xurl}
\usepackage[colorlinks=true,allcolors=black,bookmarksnumbered=true]{hyperref}
\newcommand{\doi}[1]{\href{https://doi.org/#1}{doi:\nolinkurl{#1}}}
\title{\vspace{-1.2cm}{Does the Universe Have Bugs?}\\[4pt]
\large Adversarial Physics and the Configuration Frontier}
\author{\begin{minipage}{.96\textwidth}
\centering
Wenbiao Han\textsuperscript{1,2,3}\\[5pt]
{\small\linespread{1.05}\selectfont
\textsuperscript{1}Shanghai Astronomical Observatory, Chinese Academy of Sciences, Shanghai, 200030, China\endgraf
\textsuperscript{2}School of Fundamental Physics and Mathematical Sciences, Hangzhou Institute for Advanced Study, University of Chinese Academy of Sciences, Hangzhou 310024, China\endgraf
\textsuperscript{3}School of Astronomy and Space Science, University of Chinese Academy of Sciences, Beijing 100049, China\endgraf
\vspace{4pt}
Email: \href{mailto:wbhan@shao.ac.cn}{\nolinkurl{wbhan@shao.ac.cn}};
\href{mailto:wenbiaohan@gmail.com}{\nolinkurl{wenbiaohan@gmail.com}}\endgraf}
\end{minipage}}
\date{\small 15 September 2026}
\hypersetup{pdftitle={Does the Universe Have Bugs?},pdfauthor={Wenbiao Han},pdfsubject={Adversarial physics; causal fuzzer; writable vacuum}}

\begin{document}
\maketitle
\begin{abstract}
Physics has traditionally advanced through two interrelated activities: inferring the laws of nature from observations and testing theoretical predictions with increasing precision, at higher energies, or under increasingly extreme conditions. This Perspective proposes a new research paradigm---\emph{adversarial physics}---that subjects physical limits regarded as insurmountable to active stress testing. It asks not only whether a theory correctly explains and predicts phenomena, but also whether a specially designed yet physically realizable configuration could circumvent, weaken, or even qualitatively alter the operational limits posited by that theory.

We introduce the heuristic concept of a ``\emph{cosmic bug}'' to denote an \emph{exploitable physical anomaly}: a reproducibly realizable physical configuration that grants an observer new capabilities previously considered outside the set of achievable physical operations. This concept requires neither a simulation hypothesis nor an external designer. The proposed paradigm currently encompasses at least three levels: (1) actively constructing complex systems under established physical laws to seek and exploit unforeseen loopholes in specific configurations and combinations of operations; (2) actively seeking, testing, and exploiting anomalous mechanisms beyond the explanatory scope of currently known theories, prompting revisions to our understanding of physical laws or the discovery of new physics; and (3) exploring how loopholes might be exploited to control or modify local physical rules. We further argue that traditional searches centered on the energy, intensity, and precision frontiers may have comparatively neglected another dimension---the \emph{configuration frontier}: the vast space of physical configurations comprising complex temporal sequences, phase correlations, coherent and entangled states, feedback protocols, topological structures, multiscale organization, and engineered spacetime or vacuum conditions.

To explore this frontier systematically, physical prohibitions can be decomposed into explicit assumptions, physically realizable input configurations can be generated and adjusted, feedback from outcomes can guide subsequent searches, and candidate loopholes can be experimentally tested for reproducibility and exploitability. Artificial intelligence provides an important tool for exploring this high-dimensional configuration space. This framework is related to constructor theory, resource theories, model falsification, anomaly detection, and automated experimental design, but differs in its research objective: the target of our attack is not a particular theoretical parameter, but the robustness of the boundary between the possible and the impossible itself. Even if no exploitable loopholes are ultimately found, systematic search and validation can establish progressively stronger ``\emph{no-bug bounds}'' under explicitly specified conditions. The ultimate aim is not merely to map nature's capability boundaries, but to discover and exploit possible loopholes in the laws governing it---including physically realizable routes to controlling or modifying local physical rules.
\end{abstract}

\section{Introduction: motivations and aims of adversarial physics}
Since the emergence of modern science, physics has achieved remarkable success in humanity's efforts to understand nature and transform the world. From Newtonian mechanics, which unified celestial and terrestrial motion, through electromagnetic theory, which unified electricity, magnetism and light, to quantum theory and relativity, which reshaped our understanding of matter, energy and spacetime, physics has progressively developed a quantitative description spanning microscopic particles and the macroscopic Universe. The Standard Model of particle physics and general relativity have withstood extensive experimental and observational tests. The discovery of the Higgs boson and the direct detection of gravitational waves further demonstrate the power of the interplay between theoretical prediction and precision experiment \cite{atlas2012intro,ligo2016intro}. These advances in understanding have also continually become practical capabilities: technologies such as semiconductors, lasers, nuclear energy and precision timekeeping have enabled humanity to organize matter, transmit information and use energy in unprecedented ways.

Yet this highly successful body of knowledge still faces profound difficulties. The physical nature of dark matter, the mechanism behind the accelerating expansion of the Universe, and the unification of gravity with quantum theory remain to be understood more fully. Existing theories accurately describe a vast range of known phenomena but have yet to provide a complete account of nature. At the same time, a considerable gap separates theoretical understanding from practically realizable capabilities. Even when a goal is compatible with known laws, we may not know how to construct a physical system that achieves it. The challenges of confinement, stability, materials and energy balance in practical fusion energy illustrate this gap. The former class of difficulties concerns the adequacy of our physical descriptions; the latter concerns how to obtain a target capability from those laws \cite{cernstandardintro,nasadarkenergyintro,doefusionintro}.

These difficulties invite us to examine further how new phenomena and capabilities are discovered. Increasing energy, intensity and measurement precision, extending the scope of observations, and proposing and testing new theories continue to advance physics. Yet the space of explorable physical systems contains another dimension: how inputs and operations are organized. The same energy and material resources can be arranged into different temporal sequences, phase relationships, entanglement structures, feedback protocols and forms of multiscale organization. The success of a theory across a broad range of experimental conditions does not imply that all its consequences in such special configurations have been exhausted. The robustness of a physical restriction under familiar preparation procedures also needs to be tested across a broader range of combinations of operations. This raises a question worth investigating systematically: can the deliberate construction of special physical systems reveal exploitable mechanisms not yet reached by conventional exploration, or even expose the explanatory boundaries of existing theories?

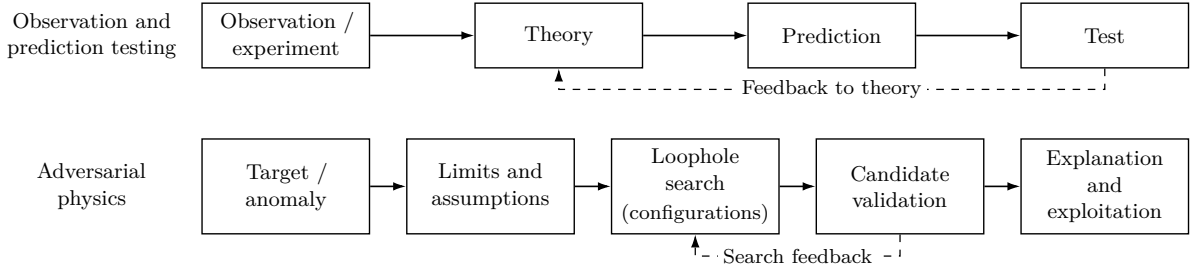
\begin{figure}[tb]
\centering
\begingroup
\linespread{1}\selectfont
\resizebox{.99\linewidth}{!}{%
\begin{tikzpicture}[x=1cm,y=1cm,
  font=\rmfamily\fontsize{8.5}{10.5}\selectfont,
  box/.style={draw=black,line width=.55pt,fill=white,
    align=center,text width=2.12cm,minimum height=.90cm,
    inner sep=3pt},
  rowlabel/.style={align=center,text width=2.70cm,font=\mdseries\fontsize{8.5}{10.5}\selectfont},
  arr/.style={-{Latex[length=1.8mm,width=1.2mm]},line width=.65pt},
  back/.style={arr,dashed,line width=.5pt},
  note/.style={fill=white,inner sep=2pt,font=\fontsize{8}{10}\selectfont}]
\node[rowlabel] at (1.30,3.15) {Observation and\\prediction testing};
\node[box] (obs) at (4.00,3.15) {Observation /\\experiment};
\node[box] (theory) at (7.80,3.15) {Theory};
\node[box] (prediction) at (11.60,3.15) {Prediction};
\node[box] (test) at (15.40,3.15) {Test};
\draw[arr] (obs)--(theory);
\draw[arr] (theory)--(prediction);
\draw[arr] (prediction)--(test);
\draw[back] (test.south)--(15.40,2.40)--(7.80,2.40)--(theory.south);
\node[note] at (11.60,2.40) {Feedback to theory};

\node[rowlabel] at (1.30,1.05) {Adversarial\\physics};
\node[box,minimum height=1.32cm] (entry) at (4.00,1.05) {Target / anomaly};
\node[box,minimum height=1.32cm] (scope) at (6.85,1.05) {Limits and\\assumptions};
\node[box,minimum height=1.32cm] (search) at (9.70,1.05) {Loophole search\\[1pt]{\fontsize{8}{10}\selectfont (configurations)}};
\node[box,minimum height=1.32cm] (verify) at (12.55,1.05) {Candidate\\validation};
\node[box,minimum height=1.32cm] (use) at (15.40,1.05) {Explanation\\and\\exploitation};
\draw[arr] (entry)--(scope);
\draw[arr] (scope)--(search);
\draw[arr] (search)--(verify);
\draw[arr] (verify)--(use);
\draw[back] (verify.south)--(12.55,.08)--(9.70,.08)--(search.south);
\node[note] at (11.125,.08) {Search feedback};
\end{tikzpicture}
}
\endgroup
\caption{{Research paths centered on prediction testing and the search for exploitable loopholes.}
The upper row summarizes the feedback between observation, theory and prediction testing. The lower row starts from a target operation or an anomalous observation, specifies a restriction and its assumptions, and uses configuration search to identify candidate loopholes for independent validation. Validation includes reproducibility, exploitability and resource accounting; only candidates that pass these checks proceed to explanation and exploitation. Numerical simulations support screening and analysis, whereas physical anomalies beyond existing theories require experimental validation. Searches that find no candidate or fail validation contribute to the search record; no-bug bounds require explicit conditions and supporting arguments. The three exploratory levels concern exploitable loopholes under established laws, anomalies beyond existing theories and new physics, and the control or modification of local physical rules. The third is an exploratory goal, and the levels are not mandatory sequential stages. The two paths represent research emphases and can intersect and iterate in practice.}
\label{fig:workflow}
\end{figure}

We therefore propose a new research paradigm---\emph{adversarial physics}. It starts from an explicit physical restriction or a target operation that cannot yet be realized, and actively seeks the physical configurations most likely to make that restriction fail. Researchers must specify the assumptions underlying the restriction, the resources and operations that are permitted, and the experimental outcomes that would constitute a genuine breakthrough. They then construct, adjust and test candidate systems in a search for anomalous mechanisms that can be reproducibly triggered and exploited. Observed anomalies can also provide starting points, motivating further intervention experiments to determine whether an anomaly can be turned into a controllable physical operation. We call this exploration of structures, correlations and combinations of operations the \emph{configuration frontier}. Artificial intelligence provides an important tool for exploring this high-dimensional configuration space. Figure~\ref{fig:workflow} summarizes the basic workflow of adversarial physics and its relationship to research centered on observation and the testing of predictions.

The concept of a ``\emph{cosmic bug}'' provides a heuristic for this research objective. In complex systems, even when the underlying rules are explicit and every operation follows them, special combinations of inputs can produce behavior that the designers did not anticipate and that an attacker can exploit. Applying this idea to physics requires neither a designer of the Universe nor a simulation hypothesis. Here, ``unanticipated'' refers to our current physical understanding and the judgments about capabilities that it supports. ``Exploitation'' requires an observer to use concrete interventions to reproducibly obtain an operation previously considered inaccessible. A loophole search must therefore state explicitly which restriction it challenges and what capability would be gained if it succeeded.

The proposed paradigm currently encompasses at least three levels. {The first level seeks loopholes under established physical laws:} actively constructing complex systems to explore whether special configurations and combinations of operations can circumvent previously assumed restrictions and reveal exploitable mechanisms that have not been fully understood. This may open new routes in materials design, energy conversion and the control of complex systems, or clarify why a particular capability remains inaccessible. {The second level seeks loopholes beyond the explanatory scope of existing theories:} deliberately producing or investigating anomalies, testing whether they can yield reproducible and exploitable physical processes, and thereby revising our understanding of physical laws or discovering new physics. {The third level explores whether loopholes could be exploited to control or modify local physical rules:} asking further whether physical interventions within the Universe could make relations treated as fixed in current fundamental descriptions into objects of control. At present, the third level is a more exploratory goal and must be clearly distinguished from known control over states, media and boundary conditions. These three levels extend the program from discovering capabilities within established laws to challenging existing laws, while retaining the question of whether the rules themselves might be controllable.

This research orientation has clear connections to existing work. Constructor theory places possible and impossible physical transformations at the center of theoretical description \cite{deutsch2012,deutschmarletto2015}; resource theories study the transformations achievable under specified resources and operations \cite{brandao2013,horodecki2013}; and automated quantum-experiment design demonstrates the ability of computational searches to discover counterintuitive configurations \cite{krenn2016}. Adversarial physics aims to organize these ideas and tools into a research strategy directed toward loophole discovery: restrictions serve as targets, physical configurations as means of intervention, and reproducible, controllable and exploitable outcomes as evaluation criteria. Particularly for the second level, research must seek evidence of deviations from existing predictions and further test whether those deviations can yield new operational capabilities, connecting anomaly identification, mechanism exploration and the discovery of new physics.

The following sections first clarify the operational meaning of a cosmic bug and the evidence required to establish one, then discuss the configuration frontier and heuristic ideas for adversarial searches. Historical examples and representative questions illustrate possible research directions. Successful searches may reveal overlooked resources, open new physical capabilities or provide experimental evidence beyond existing theories. Unsuccessful searches can establish stronger no-bug bounds under explicitly specified conditions. This Perspective seeks to advance an organized, active exploration of natural laws and the capabilities they permit: using deliberately constructed physical processes to discover and exploit possibilities that remain beyond our present understanding.

\section{What is a cosmic bug?}\label{sec:formal}
\subsection{The concept of a cosmic bug and its three levels}
We define a ``cosmic bug'' as an exploitable physical mechanism: through specific, physically realizable system configurations or combinations of operations, an observer can reproducibly and controllably realize physical processes previously considered unattainable, thereby acquiring new operational capabilities. A ``loophole'' here is relative to an explicit physical description, assumptions and resource conditions; its mechanism may lie within established laws or may require revising or extending the existing physical description.

Table~\ref{tab:levels} summarizes the three levels of this research program. The first and second levels distinguish whether a mechanism can be explained by existing theories; the third further asks whether local physical rules can become controllable.

\begin{table}[!htbp]
\centering\small
\begin{tabularx}{\linewidth}{@{}p{0.065\linewidth}p{0.235\linewidth}X@{}}
\toprule
Level & Target & Evidential interpretation\\
\midrule
I & Exploitable loopholes under established laws & Special configurations or combinations of operations circumvent a previously assumed restriction; the mechanism remains explainable by established physical laws.\\
II & Loopholes beyond existing theories & A reproducible, exploitable anomaly remains beyond the explanatory scope of existing theories after independent validation and checks of resources, approximations and known mechanisms, constituting a candidate for new physics.\\
III & Control or modification of local physical rules & A concrete physical intervention makes a local relation treated as fixed in the current fundamental description controllable; this is an exploratory goal.\\
\bottomrule
\end{tabularx}
\caption{Three levels of research on cosmic bugs. Each level is relative to an explicit reference theory and capability restriction. The levels express an expansion of research goals, rather than mutually exclusive categories or stages that must be traversed. The third level is exploratory; if realized, its mechanism must still be assessed for compatibility with existing theories.}
\label{tab:levels}
\end{table}

{The first level: finding and exploiting loopholes within established physical laws.} This level assumes that established physical laws hold and asks whether complex systems contain unforeseen exploitable mechanisms. Even if each constituent process follows known rules, we may not have exhausted the operations that combinations of these processes can realize. An adversarial search therefore starts from a capability considered unattainable and deliberately constructs special system structures, input sequences or feedback protocols to try to circumvent previously assumed restrictions. What is overcome here is an earlier judgment about the system's achievable capabilities. That judgment may reflect an incomplete understanding of complex configurations and combinations of operations, or arise from simplified models or incomplete resource descriptions. The discovered mechanism remains explainable by established laws. A {domain loophole}, in which a prohibition has been extended beyond its assumptions, belongs to this level, as do unforeseen exploitable mechanisms in complex combinations of operations. For example, one may envisage realizing usable room-temperature superconductivity through special material configurations, although this goal is not itself subject to a universal prohibition in existing theory \cite{drozdov2019}.

{The second level: finding and exploiting loopholes beyond the explanatory scope of currently known physical theories.} This level explores the explanatory boundaries of existing theories, deliberately constructing physical processes that may expose their limitations or starting from observed anomalies to ask whether they reveal mechanisms absent from the existing description. If an anomaly has undergone independent validation and still requires going beyond currently known theories after checks of resources, approximations and known physical mechanisms, it constitutes candidate evidence for revising our physical understanding or discovering new physics. Adversarial physics asks a further question: can concrete interventions repeatedly trigger this mechanism and turn it into a new operational capability? The existence of an anomaly provides a clue for a search; reproducible, controllable exploitation is what makes it satisfy this paper's definition of a loophole. For example, the dark-matter problem may serve as an anomalous clue in the search for new physics \cite{cernstandardintro}, although whether it belongs to this level depends on the eventual mechanism.

{The third level: exploring the exploitation of loopholes to control or modify local physical rules.} This level further asks whether a physical mechanism could make relations governing local processes controllable, even when they are treated as fixed in the current fundamental description. Its goal extends from acquiring a new operational capability to actively intervening in the rules governing local processes and using that intervention to change which physical operations are achievable. Establishing such control requires specifying the relation subject to intervention and distinguishing controllable changes in rules from responses to changes in states, media or boundary conditions under existing laws. At present, this remains an exploratory goal whose physical realizability has yet to be established. For example, exploring local control over a gravitational relation previously treated as fixed can provide a hypothetical illustration of this goal.

These examples serve only to aid understanding; they neither define the levels nor represent confirmed cosmic bugs. To turn the three levels of exploration into testable research, we must make explicit the basis for judging a capability ``previously unattainable'': which operation is sought, which physical description and assumptions support the original restriction, which resources may be used, and how success is to be judged. These conditions allow us to distinguish new capabilities within established laws from anomalies beyond existing theories, and provide a basis for assessing claims of rule control. The next subsection therefore introduces the task and reference description as an explicit starting point for a loophole search.

\subsection{Specify the task and the reference description}
Specifically, a search should specify the target operation, the theoretical description used for comparison, and the permitted resources and success criteria. This set of conditions can be written as
\begin{equation}
 \mathfrak C=(T,\mathcal H,\mathcal R,\Pi,\mathsf T,\mathsf V,\varepsilon).
 \label{eq:contract}
\end{equation}
Here $T$ is the reference theory, $\mathcal H$ its assumptions and approximation regime, $\mathcal R$ the available resources, $\Pi$ the candidate preparation protocols, $\mathsf T$ the target task, $\mathsf V$ its success witness, and $\varepsilon$ the stated tolerance. Apparatus, ancillary systems, communication, postselection and reset belong in the resource description. Changing the input ensemble, deadline or accepted failure outcomes changes the task.

Preparation instructions must be distinguished from the operation they are predicted to implement. Let $\mathcal E_T(\pi)$ be the predicted map and $\mathcal K_T(\mathfrak C)$ the reference reachable set. An experimental candidate can satisfy
\begin{equation}
 \mathcal E_T(\pi)\in\mathcal K_T(\mathfrak C),\qquad
 d_{\mathsf T}\!\left(\mathcal E_{\rm lab}(\pi),\mathcal K_T(\mathfrak C)\right)>\varepsilon,
 \label{eq:empiricalgap}
\end{equation}
where the distance refers to observable input--output statistics. This avoids claiming that an operation both belongs and does not belong to the same exact closure. What fails can be a preparation model, an omitted resource, an approximation, or the physical theory. A theoretical loophole may instead show that the old domain excluded a realizable resource; it does not contradict a theorem on its original domain.

\subsection{Evidential requirements and exploratory boundaries}
A first-level loophole need not produce an experimental deviation from the complete predictions of established theory; it may consist of a configuration allowed by existing laws that circumvents a previously assumed restriction. Equation~\eqref{eq:empiricalgap} describes an experimental deviation relative to the reference description. Only if the deviation remains unexplained by existing theories after checks of the preparation, resources and approximations does it constitute a second-level candidate.

A discovery at any level may enable further operations, but each additional capability still requires independent validation. An anomalous clue can initiate a search; establishing an exploitable loophole requires a concrete intervention, reproducibility and a measurable use. An unsuccessful search can strengthen support for a restriction, but such a valuable negative result cannot replace the definition of a cosmic bug.

The third level remains an open question. A test must specify which local relation is treated as fixed and what intervention could control it. Geometry is already dynamical in general relativity, and effective couplings can depend on background fields. Thus known responses to changes in matter distributions, system states or boundary conditions do not by themselves establish rule control. For a dimensionless relation $r$, one may test
\begin{equation}
 H_0^{\rm rule}:\quad |\Delta r_{\rm residual}(\pi)|\leq\varepsilon,
 \qquad \pi\in\mathcal D,
 \label{eq:rulenull}
\end{equation}
where known responses have been accounted for over the specified domain $\mathcal D$, and instrumental effects must be checked independently. A persistent, independently controllable violation would provide candidate evidence for rule control; a deeper theory might explain it through new fields or interactions. Obtaining such control would not establish that an unknown final law had been violated.

\section{Methodological ideas: configuration search and loophole validation}\label{sec:frontier}
Section~2 defined cosmic bugs and the evidence required to establish them; this section considers how these goals might be formulated as search problems. {This Perspective aims to propose a heuristic research approach. The configuration representations, search strategies and order-of-magnitude estimates below are illustrative; practically feasible methods, their domains of applicability, search efficiency and implementation requirements remain to be studied systematically in specific physical systems.} These ideas can provide starting points for research design, but their effectiveness still requires assessment through theoretical analysis, numerical benchmarks and experiments.

\subsection{The configuration frontier: construction ideas and search-space size}
\begin{figure}[tb]
\centering
\resizebox{.99\linewidth}{!}{%
\begin{tikzpicture}[x=1cm,y=1cm,
 font=\rmfamily\fontsize{11}{13}\selectfont,
 arr/.style={-{Latex[length=2.5mm]},line width=.85pt}]
\draw[arr] (0,0)--(12.0,0);
\draw[arr] (0,0)--(0,5.7);
\node[anchor=south west] at (0,5.82) {Configuration complexity $K(X)$};
\node[anchor=north east] at (12.0,-.22) {Energy / intensity / scale / precision};
\fill[black!9] (.3,.35) rectangle (11.4,1.35);
\node[align=center,text width=10.8cm] at (5.85,.85) {traditional energy / intensity / scale / precision frontiers};
\fill[black!16] (.5,1.82) rectangle (4.0,4.98);
\node[align=center,text width=3.15cm] at (2.25,3.4) {programmable\\quantum and\\control platforms};
\draw[dashed,line width=.9pt] (4.72,1.73) rectangle (11.4,5.3);
\node[font=\mdseries,align=center,text width=6.2cm] at (8.06,4.66) {configuration frontier};
\node[align=center,text width=6.1cm] at (8.06,3.34) {structured, correlated,\\feedback-controlled,\\topological and multiscale inputs};
\draw[arr,line width=1.3pt] (4.12,2.0)--(5.28,2.77);
\end{tikzpicture}
}
\caption{{The configuration frontier complements traditional frontiers in energy, intensity, scale and precision.} Input organization---phase, temporal order, correlations, topology and composition---can be explored at specified resources and measurement conditions. The horizontal axis schematically groups distinct dimensions of exploration, including access to larger and smaller scales; it is not a common quantitative scale. The regions are schematic and overlap in practice; $K(X)$ is a problem-dependent structural descriptor, not a universal measure or a claim that greater complexity guarantees new physics.}\label{fig:frontier}
\end{figure}
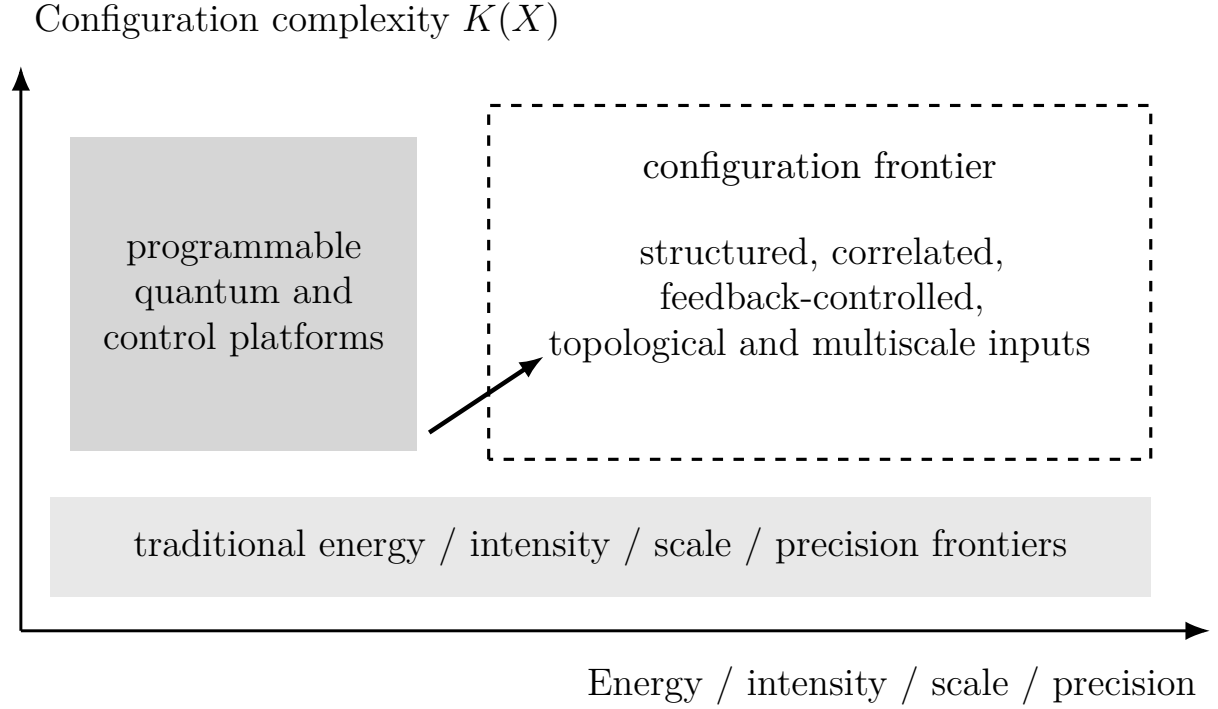

Increasing energy and intensity, extending the scales accessible to observation, and improving measurement precision are all important routes to exploring physical phenomena. Exploration across scales includes both the larger spatial extents and structural scales covered by cosmological surveys and the probing of physical processes at smaller scales. These routes can also be combined with different ways of organizing physical inputs. The {configuration frontier} concerns how temporal order, phase, correlations, topology and feedback are organized under explicit resource conditions (Fig.~\ref{fig:frontier}). It assumes neither a universal complexity scale nor that more complex inputs are more likely to yield valuable results.

One construction scheme worth investigating starts from a {library of elementary operations and rules for combining them}. A physical platform, initial state, measurement scheme and resource budget would first be specified; implementable preparation, driving, coupling, measurement and feedback operations would then be listed, together with their parameter ranges, resolution, interfaces and resource costs. Candidates could be represented as finite sequences of operations or parameterized connection graphs, with composition rules specifying serial and parallel operations, conditional triggering and the permitted total duration. The operation library should constrain the interventions that can actually be implemented, without requiring in advance that experimental outcomes conform to the predictions of the theory under test.

Configuration expansion could start from short protocols, progressively changing order, correlations or connections while recording the cost of each extension. The physical basis for selecting these changes could be a specific assumption underlying a restriction: for example, comparing different orders of noncommuting operations, coherent inputs with randomized-phase controls, open-loop control with conditional feedback that includes resource accounting, or combinations of processes at different temporal and spatial scales. For promising candidates, operations could also be removed one at a time to identify simpler combinations that sustain the effect and explore their reuse. These are design guidelines to be tested; they do not guarantee a general advantage for any class of configurations.

Existing studies provide reference points for some of these construction ideas. MELVIN assembles apparatus from a library of experimentally available optical elements, evaluates the results against a target, and adds simplified successful combinations to the library for subsequent searches \cite{krenn2016}. Coherent control and optimization of Schwinger pair production also illustrate uses of pulse shaping and inverse design \cite{judson1992,kohlfurst2013,hebenstreit2014,hebenstreit2016}. These studies support the feasibility of systematic searches over specific configurations, but do not establish the existence of cosmic bugs or the efficiency with which they might be found.

{Search-space size must be estimated relative to a specific representation and resolution.} Suppose that a control process contains $L$ time slots, each selecting from $m$ calibrated and distinguishable operations. The nominal number of encoded protocols of fixed length is then
\[
 N_{\mathrm{code}}=m^L.
\]
For example, suppose that each time slot in a fixed pulse template allows a choice among four phases, $0,\pi/2,\pi,3\pi/2$. For $L=10,20,30$, the numbers of encoded sequences are approximately $1.05\times10^6$, $1.10\times10^{12}$ and $1.15\times10^{18}$, respectively. If a cost of $1\,\mathrm{ms}$ per candidate is assumed solely to illustrate the rate of growth, fully serial exhaustive evaluation would take approximately 17.5 minutes, 34.8 years and $3.65\times10^7$ years, respectively; these are not performance estimates for any experimental platform. For $d$ continuous parameters, if parameter $i$ is discretized into $q_i$ distinguishable settings, the corresponding nominal grid size is $\prod_{i=1}^{d}q_i$.

These numbers count only candidates within a specified representation; they cannot be interpreted as the total number of configurations in the Universe or the number of distinct physical effects. Unimplementable combinations, sequences exceeding the budget, and candidates equivalent under symmetries or actual resolution must still be identified. If a symmetry is itself among the assumptions under test, it should not be used to merge candidates in advance. Parameter resolution should be supported by control and measurement capabilities; within the same finite encoding space, the number of feasible and distinguishable configurations cannot exceed the nominal number of encoded candidates. Exponential growth in candidate counts alone does not determine the actual difficulty of searching for a particular task.

\subsection{Adversarial search: strategy selection and resource budgets}\label{sec:fuzzer}
A search can start from a target capability or from an observed anomaly. The former can draw on inverse design, first specifying a desired operation and then seeking preparations that might realize it; the latter can first investigate whether an anomaly can be reproducibly triggered and controlled, and then explore its uses. Both paths ultimately require the task, reference description and resource conditions introduced in Section~2. For a target-driven search, the relevant restriction can be decomposed schematically as
\begin{equation}
 A\wedge B\wedge C\ \Longrightarrow\ \neg\mathsf T.
 \label{eq:assumptions}
\end{equation}
Here $A$, $B$ and $C$ schematically denote the premises supporting the restriction; $\wedge$ means ``and'', indicating that these premises hold jointly; $\Longrightarrow$ denotes logical implication; and $\neg$ denotes negation. Whereas $\mathsf T$ denotes the target task in Section~2.2, in this equation the same symbol abbreviates the proposition that the task is realizable under the specified conditions. The expression therefore states that if $A$, $B$ and $C$ all hold, the target task cannot be realized.
Each premise can become a question for further study: is it a mathematical identity, an experimentally supported principle, a resource limitation or an approximation? A proposal to relax a premise should also specify a physical intervention that might realize that relaxation and its resource cost.

One search arrangement that could be tested would begin with a baseline of short protocols or simple configurations, measuring noise, drift and the cost of a single evaluation before progressively adding degrees of freedom. For a small number of continuous parameters whose evaluation is expensive, surrogate models or Bayesian optimization could be considered for choosing the next settings; for discrete operations and connection structures, candidates could be modified by insertion, deletion, exchange and recombination; where a trustworthy differentiable model is available, gradient-based or optimal-control methods could be investigated. Algorithm selection depends on the structure and noise of the task, and performance should be compared with random search or simple scans under the same resource and evaluation budgets. Artificial intelligence provides an important tool for exploring this high-dimensional configuration space, but its specific benefits still require validation in each case.

The process of continually generating, mutating and testing inputs can also draw on the idea of fuzzing in software engineering. When applied to actively stress-testing physical restrictions, we call it {physical-law fuzzing}: preparation and operation protocols are directly adjusted, while the test concerns whether the relevant restrictions continue to hold under these interventions. Anomaly-driven, open-ended searches can provide initial clues; promising responses then need to be formulated as explicit intervention-and-readout tasks. Feedback can guide the selection of subsequent candidates, but exploratory sampling should be retained to assess whether model preferences overlook other directions.

Evaluation objectives must correspond to the different levels. The first level concerns whether a capability previously considered unattainable can be obtained under the agreed resources; its outcome may conform to the full predictions of established theory. The second level additionally concerns experimental deviations that remain unexplained by existing theories after checks, and tests their exploitability. The third level requires dedicated tests of controllable changes in specified local relations. The relevant criteria are those of Section~2. New responses obtained by optimization must still be compared with known mechanisms; for example, matching energy or power spectra does not exclude known nonlinear responses caused by different spectral phases.

{Nominal search-space size must be distinguished from the number of evaluations actually performed.} The cold-atom experiment of Wigley and colleagues provides a concrete reference: for their 16-parameter scheme, taking 10 settings per parameter gives a coarse grid of $10^{16}$ points; in one reported run, the machine-learning method completed that optimization task with approximately 10 evaluations following 20 initialization evaluations \cite{wigley2016}. This illustrates that, for particular response structures, the required number of evaluations can be much smaller than the grid size; the result cannot be extrapolated into an estimate of loophole-search efficiency.

The practical budget also depends on the cost of validating each candidate. If $Q$ candidates are evaluated, each repeated $r$ times, and the complete experimental cycle takes $t_{\mathrm{cycle}}$, an illustrative accounting is
\[
 T_{\mathrm{exp}}\approx Qr\,t_{\mathrm{cycle}}
 +T_{\mathrm{calibration}}+T_{\mathrm{analysis}}.
\]
The cycle should include preparation, switching, evolution, readout and reset, with additional calibration and analysis accounted for separately; experiments with varying durations can be budgeted using their measured average costs. The $r$ repetitions of the same candidate estimate its response and uncertainty; they do not count as $r$ independent search candidates. For example, under the assumptions $Q=1000$, $r=100$ and $t_{\mathrm{cycle}}=1\,\mathrm{s}$, the repeated experiments alone would take approximately 27.8 hours. This is an illustrative budget, not a recommended universal search scale; the cost of numerical simulation must be estimated separately according to system size, representation and precision.

If candidates are sampled independently from a fixed distribution, with probability $p$ of drawing a qualifying candidate on each trial, the probability of at least one hit in $Q$ draws is
\[
 P_{\mathrm{hit}}=1-(1-p)^Q.
\]
Here $p$ is the hit probability under the specified search distribution, not the probability that a loophole exists in nature; it is currently unknown, so this expression cannot predict a discovery time. Feedback-guided searches generally change the sampling distribution, and a constant-$p$ model cannot be directly applied to them either. If a problem lacks usable structure or additional information, searching can remain extremely difficult. Identifying which physical problems permit effective reductions in search effort is a central question for subsequent methodological research.

\subsection{Candidate validation and limits of applicability}
Research following the appearance of a candidate could include repeated measurements and resource checks, checks for numerical and instrumental errors, removal of individual key operations and control comparisons, and independent validation and exploitability tests after the protocol has been fixed. The order and cost of these steps depend on the platform; results selected from large numbers of candidates particularly require validation with independent data to control selection effects. Unsuccessful results should record the search range, sensitivity and resource conditions to inform revisions of the reference description or the design of subsequent searches; finding no candidate in a finite search is not sufficient by itself to establish a universal prohibition.

Three kinds of evidence must still be distinguished. {Within-theory calculations} explore consequences or check implementations; {calculations with relaxed assumptions} explicitly change premises and record their costs; {experiments} compare independently implemented interventions with reference predictions. If an exact theorem fixes a witness to zero over its entire domain, optimization within the same model cannot circumvent the proof. Useful next questions concern the model's physical completeness, its approximations or an explicitly changed premise. Particularly for the second level, models can help propose candidates, but physical evidence beyond existing theories still requires experimental testing.

Realizability checks concern instruments and resources, rather than requiring outcomes to obey the rule being attacked. A reproducible experimental deviation cannot be rejected merely because it contradicts the reference theory. Conversely, directly adding the target communication channel as an elementary operation does not constitute its discovery. No anomaly score can compensate for such an undeclared resource.

Software testing has produced fuzzing frameworks for physics simulation engines. For example, PHYFU mutates initial states and uses test feedback to arrange subsequent inputs, checking for logic errors in forward simulation and backward gradient computation \cite{phyfu2023}. It illustrates how input mutation, feedback-guided search and dedicated criteria can be organized into a testing procedure; its findings are defects in simulation software and cannot serve as evidence for loopholes in nature. Its specific criteria depend on assumed physical conditions and would need to be re-established and tested when transferred to experimental systems. Apparent loopholes encountered in simulations may also arise from software, giving analytic controls, independent implementations and validation after fixing the protocol practical importance.

The discussion above offers a set of heuristic starting points for configuration representation, search budgeting and candidate checks. Which composition rules are effective on different platforms, which feedback can improve search efficiency, and what budgets suffice to yield meaningful constraints remain to be studied systematically.

\section{Representative questions and research directions}\label{sec:directions}
Historical research has repeatedly gained new insights by actively challenging physical restrictions. Drawing on these precedents, we outline several questions worth exploring, with an emphasis on discovering capabilities within established laws and exploiting anomalies beyond existing theories, while retaining the longer-term question of whether local rules might be controllable. These questions illustrate research directions; their implementation and physical feasibility remain to be studied.

\subsection{Historical inspiration: actively challenging physical restrictions}\label{sec:history}
\paragraph{Maxwell's demon: from an apparent violation to information as a resource.}
Maxwell's demon envisages obtaining microscopic information and applying conditional feedback to extract work cyclically from an equilibrium reservoir. The challenge is concrete: it goes beyond questioning the second law to attempt to construct a protocol that uses measurement outcomes to select subsequent operations. In a standard isothermal feedback setting, the work extractable from the controlled subsystem satisfies
\begin{equation}
 \langle W_{\rm ext}\rangle\leq-\Delta F+k_{\rm B}T I,
 \label{eq:feedback}
\end{equation}
where $\Delta F$ is the subsystem's free-energy change, $T$ is the reservoir temperature, and the information $I$ is measured in nats and corresponds to a specified measurement model \cite{sagawa2008,sagawa2010}. Feedback experiments show that information can be converted into a usable advantage in work extraction \cite{toyabe2010}. Completing the entire cycle, however, also requires restoring the memory and controller to usable states. For an ideal degenerate memory containing an unbiased bit with no useful side information, the usual minimum work for isothermal reset is $k_{\rm B}T\ln2$; its conditions of applicability cannot be extended to arbitrary measurement or erasure processes \cite{bennett2003}.

This example yields both new information-assisted capabilities and a reformulation of the original restriction: accounting only for subsystem work extraction omits resources involved in measurement, feedback and reset. Actively constructing processes that appear to violate a restriction thus helped deepen understanding of the relationship between information and thermodynamics. It provides inspiration for first-level exploration, rather than evidence that the fully stated second law has been overturned.

\paragraph{Black-hole extremality: seeking and testing weaknesses in a boundary.}
In four-dimensional classical Einstein--Maxwell theory, using geometrized units $G=c=4\pi\epsilon_0=1$, the mass $M$, charge $Q$ and angular momentum $J$ of a stationary, asymptotically flat Kerr--Newman black hole satisfy
\begin{equation}
 \mathcal D=M^4-M^2Q^2-J^2\geq0.
 \label{eq:extremality}
\end{equation}
A concrete challenge is to select matter that can be absorbed by the black hole and delivers sufficiently large charge or angular momentum relative to its energy, in an attempt to cross this boundary. The condition $\mathcal D<0$ is the corresponding diagnostic only when the final stationary-family description applies; it cannot directly replace an examination of horizons in a dynamical spacetime.

Wald's thought experiment for extremal black holes found that test bodies capable of threatening the boundary could not be captured in the required way \cite{wald1974}. Later near-extremal test-body analyses identified apparent windows for overcharging or overspinning \cite{hubeny1999,jacobson2009}, shifting the question to whether these windows could survive a more complete calculation. The Sorce--Wald analysis included the necessary second-order effects under its perturbative, energy and stability assumptions and closed the corresponding windows \cite{sorcewald2017}. This sequence illustrates how to select challenging inputs, identify gaps in an approximation, and then test whether corrections at the same order restore the restriction. The conclusion remains conditional; further investigations of other matter, quantum effects or different assumptions require the applicable description to be established anew. Both historical examples demonstrate the research value of concrete challenges, but neither directly establishes a confirmed cosmic bug.

\subsection{Some possible explorations: complex configurations and anomalous mechanisms}\label{sec:causal}
The first level could start from target capabilities not yet realized, investigating whether particular structures and combinations of operations can circumvent a previously assumed restriction. The second level further tests whether anomalies exceed existing theories and examines whether they can be controlled and exploited. Two possible examples are given below.

\paragraph{Strong-field responses: testing configuration searches through known mechanisms.}
Existing studies of Schwinger pair production provide references for pulse optimization and inverse design \cite{kohlfurst2013,hebenstreit2014,hebenstreit2016}. One starting point worth investigating is to specify input energy, peak field, duration, bandwidth and geometry, then compare particle yields for different phases, temporal orders and pulse combinations. This could test whether changes in configuration yield responses difficult to obtain with conventional inputs, and whether any advantage arises from unaccounted resources or known nonlinear mechanisms. A higher yield does not automatically constitute a loophole; only relative to an explicit capability restriction can one assess whether a new exploitable route has been found. A response that remains unexplained by known theories after independent checks would then raise a second-level question.

\paragraph{Causality searches: specifying the target and calculable controls.}
A stronger question is whether a sender's independently chosen setting can produce a usable signal in a region spacelike separated in the actual spacetime. Let Alice choose $a\in\{0,1\}$ independently of the prepared state and shared information, and let Bob record $b$ before a fixed deadline. All setting-dependent interventions must be spacelike separated from Bob's readout region, and all outcomes, including no-clicks, must be retained. The target can be expressed as a difference in raw statistics:
\begin{equation}
 S_{\rm raw}(\pi)=\frac12\sum_b
 \left|P_\pi(b\mid\operatorname{do}(a=0))
       -P_\pi(b\mid\operatorname{do}(a=1))\right|.
 \label{eq:signal}
\end{equation}
Here $\operatorname{do}$ denotes an active choice; with equal priors, the optimal single-shot decoding success probability is $(1+S_{\rm raw})/2$. For a fixed initial state, Alice's local, unconditional completely positive trace-preserving (CPTP) operation leaves Bob's reduced state unchanged, giving $S_{\rm raw}=0$. This analytic control cannot be circumvented through optimization within the same model; a global quantum operation is also not necessarily realizable through local operations \cite{beckman2001}.

Three simple controls make the meaning of this criterion concrete (Table~\ref{tab:calibration}). The random local operations were checked for 100 paired settings at each of $d_A=d_B=2,3,4$, using random seed \texttt{20260908}. For the Bell state $|\Phi^+\rangle=(|00\rangle+|11\rangle)/\sqrt2$, Alice applies the filter $K_a=|a\rangle\langle a|$, leaving Bob in state $|a\rangle$ within the successful subensemble, apparently enabling perfect signalling. Each filter succeeds with probability $1/2$, however; restoring failed outcomes leaves Bob in state $I_B/2$ for both settings. Bob cannot exploit the selection advantage without receiving the success flag in time. Conversely, directly permitting a cross-party CNOT interaction already supplies a communication resource.

\begin{table}[!htbp]
\centering\small
\begin{tabularx}{\linewidth}{@{}p{.25\linewidth}p{.26\linewidth}X@{}}
\toprule
Control & Result & Interpretation\\
\midrule
Random local CPTP maps & $\max S_{\rm raw}<5\times10^{-16}$ & Bob's statistics remain unchanged within floating-point error across 300 paired settings.\\
Bell-state filtering & $S_{\rm selected}=1$; $S_{\rm raw}=0$ & Conditional discrimination requires a remote success flag and does not provide local spacelike communication.\\
Explicit cross-party CNOT & $S_{\rm raw}=1$ & An interaction capable of conveying information has been permitted; this is not a local spacelike protocol.\\
\bottomrule
\end{tabularx}
\caption{Three predetermined controls for a causality search. The numerical values correspond to this implementation check of the fixed cases; the last digits of the residuals depend on the floating-point environment. The cases were not discovered through adaptive search and provide neither evidence for new physics nor experimental limits on faster-than-light signalling.}\label{tab:calibration}
\end{table}

\subsection{Long-term exploration: the controllability of local physical rules}\label{sec:vacuum}
In the usual quantum-field-theoretic description, the vacuum is not an absence of physical content, but the ground state for a specified background and boundary conditions. Some theories admit multiple vacua or metastable vacua whose field backgrounds and excitation properties can differ \cite{coleman1977}. Intervening in the vacuum therefore concerns not only particle production, but also whether different field backgrounds can be prepared and maintained. Here, ``writing'' borrows the operational meaning of storage: actively preparing a local state that remains readable after the drive is removed and can be erased or rewritten.

``Can the vacuum be written?'' is first a question about state preparation and retention. Existing Floquet vacuum engineering provides a reference for controlled responses under an external drive, but does not establish erasable and rewritable rule control after the drive is removed \cite{yamada2021}. The third level poses a stronger question: could such an intervention make local relations treated as fixed in the current fundamental description amenable to control? Assessing this requires distinguishing it from responses caused by changing field states under established laws.

A double-well scalar field can serve as a simplified model for studying this state-preparation problem. With $\hbar=c=1$, one can study
\begin{equation}
 \mathcal L=\frac12\partial_\mu\phi\partial^\mu\phi
 -\frac\lambda4(\phi^2-v^2)^2+J(\vec{x},t)\phi,
 \label{eq:scalar}
\end{equation}
where $\lambda>0$ and $v>0$ specify the double-well potential. A drive $J$ of finite duration and spatial extent could attempt to move the local field from near $+v$ to near $-v$, followed by an examination of retention time and reset costs after the drive is removed. An instantaneous reversal is not a stable record: for degenerate vacua in three spatial dimensions, the energy of a spherical thin-wall domain is approximately $4\pi\sigma R^2$, where $R$ is the domain radius and $\sigma$ is the domain-wall tension; it provides no stable minimum at finite radius \cite{coleman1977}. If charges, defects, other fields or boundaries are required to stabilize the region, they must be explicitly included as resources.

This model could be used to study the costs and trade-offs among writing, retention and rewriting; success would first establish state control within a known field theory. A further claim of rule control would require independent dimensionless probes to distinguish ordinary particles, residual fields, known states and instrumental effects; optical-clock frequency ratios provide a reference for probe design \cite{rosenband2008}. Reproducible control of local relations requires evidence beyond these known responses; a persistent signal alone is insufficient. Control of local rules remains an open question whose physical feasibility has yet to be established.

\section{Discussion and conclusion}\label{sec:conclusion}
The remarkable success of physics does not imply that we have exhausted the operational capabilities permitted by nature, or that existing theories can explain every physical process. Current difficulties have two distinct sources: a vast configuration space still separates known laws from desired capabilities, while phenomena that remain incompletely explained continue to prompt scrutiny of existing theories. Increasing energy and intensity, extending observational scales and improving measurement precision continue to advance these questions. How the structures, correlations and operations of physical systems are organised may likewise determine which phenomena can be discovered and which capabilities can be realised. Starting from an explicit capability restriction or an anomalous observation, adversarial physics actively constructs specific physical processes to explore unrecognised, exploitable mechanisms within established laws and to seek anomalies beyond the explanatory scope of existing theories. The motivation for this paradigm is to translate our understanding of nature into active exploration of its space of possibilities.

A ``cosmic bug'' here denotes an exploitable physical mechanism: a particular configuration or combination of operations enables an observer to obtain, reproducibly and controllably, a capability previously considered inaccessible. Defined relative to an explicit physical description, assumptions and resource conditions, this concept accommodates unforeseen complex mechanisms within known laws and allows us to question the explanatory limits of existing theories. The corresponding adversarial approach is guided by a target capability: it actively constructs systems and interventions most likely to challenge an existing restriction, then adjusts configurations and search directions in response to the results. Its core is to connect the identification of restrictions, the generation of candidate configurations, independent validation and tests of exploitability, progressively turning anomalous observations into physical processes that can be repeatedly triggered and studied. The discovery of phenomena, the understanding of mechanisms and the acquisition of operational capabilities can thereby inform one another within the same research process.

Advances in artificial intelligence create new conditions for this exploration. The difficulty of complex configurations lies both in the vast number of candidates and in the fact that effective structures and ways of combining them are often difficult to propose through intuition alone. AI can help generate non-intuitive configurations, explore different paths in parallel and adjust search directions using computational or experimental feedback. FunSearch combined language models with evaluation programs to discover new mathematical combinatorial constructions \cite{romeraparedes2024}. Recently released AI-assisted work on Navier--Stokes singularity construction reports finite-time singularities in the three-dimensional incompressible equations with smooth forcing, providing another point of reference for construction and proof searches under complex constraints \cite{openai2026ns}. These developments suggest that AI could become a key tool for systematic exploration of the configuration frontier, sustaining some searches previously constrained by human effort and intuition. In searches for loopholes beyond existing theories, computational models can propose and screen candidates, whereas new physical mechanisms must still be established experimentally.

The historical examples suggest that active challenges can reveal overlooked resources and capabilities, or clarify the conditions under which a restriction holds. The directions proposed here connect capability discovery in complex configurations, the exploitation of anomalies beyond existing theories, and the longer-term question of whether local physical rules can be controlled. These remain research ideas to be tested. After a loophole is found, a deeper explanation may follow. Conditional limits remain valuable when an attack fails.

The potential significance of adversarial physics, the new paradigm proposed here to search for ``cosmic bugs'', lies in offering a new way to organise research that advances fundamental understanding and practical capabilities together. Its central goal is to obtain reproducible and controllable physical operations previously considered inaccessible, through mechanisms that may lie within established laws or require revisions to the existing physical description. If realised, advances at the first level could turn mechanisms hidden within complex systems into new capabilities for organising matter, converting energy and processing information. At the second level, anomalies could develop from observations awaiting explanation into phenomena amenable to active intervention, providing experimental routes to the discovery of new physics. At the third level, the question of whether local rules can be controlled extends this inquiry to the deeper relationship between natural laws and the operational capabilities of observers within the Universe. Adversarial physics thus aims to sustain a cycle of discovery: new operational capabilities expose new physical questions, while new physical understanding expands the possibilities that can be actively explored and exploited.

\begin{quote}
\emph{The question is not only what laws govern the Universe, but whether deliberately constructed physical processes can reveal, exploit, or control something those laws currently declare inaccessible.}
\end{quote}

\end{document}